\documentclass[conference]{IEEEtran}
\IEEEoverridecommandlockouts

\usepackage{cite}
\usepackage{amsmath,amssymb,amsfonts}
\usepackage{algorithmic}
\usepackage{graphicx}
\usepackage[T1]{fontenc}
\usepackage{times}
\usepackage{textcomp}
\usepackage{xcolor}
\usepackage{threeparttable}
\def\BibTeX{{\rm B\kern-.05em{\sc i\kern-.025em b}\kern-.08em
    T\kern-.1667em\lower.7ex\hbox{E}\kern-.125emX}}
\begin{document}

\title{CADC: Crossbar-Aware Dendritic Convolution for Efficient In-memory Computing}
\author{
    \IEEEauthorblockN{
        Shuai Dong,
        Junyi Yang, 
        Ye Ke,
        Hongyang Shang,
        Arindam Basu
    }
    \IEEEauthorblockA{
        Department of Electrical Engineering, City University of Hong Kong, Hong Kong, China\\
    }
\thanks{
This work was supported by the Research Grants Council of the HK SAR, China (Project No. CityU 11212823 and HKU C7003-24Y). Any opinions, findings, conclusions, or recommendations expressed in this material do not reflect the views of the Government of the Hong Kong Special Administrative Region, the Innovation and Technology Commission, or the Innovation and Technology Fund Research Projects Assessment Panel. (Corresponding authors: Junyi Yang; Arindam Basu. e-mail: junyiyang8-c@my.cityu.edu.hk; arinbasu@cityu.edu.hk.)
}
}

\maketitle

\begin{abstract}
Convolutional neural networks (CNNs) are computationally intensive and often accelerated using crossbar-based in-memory computing (IMC) architectures. However, large convolutional layers must be partitioned across multiple crossbars, generating numerous partial sums (psums) that require additional buffer, transfer, and accumulation, thus introducing significant system-level overhead.
Inspired by dendritic computing principles from neuroscience, we propose crossbar-aware dendritic convolution (CADC), a novel approach that dramatically increases sparsity in psums by embedding a nonlinear dendritic function (zeroing negative values) directly within crossbar computations. Experimental results demonstrate that CADC significantly reduces psums, eliminating 80\% in LeNet-5 on MNIST, 54\% in ResNet-18 on CIFAR-10, 66\% in VGG-16 on CIFAR-100, and up to 88\% in spiking neural networks (SNN) on the DVS Gesture dataset. The induced sparsity from CADC provides two key benefits: (1) enabling zero-compression and zero-skipping, thus reducing buffer and transfer overhead by 29.3\%, and accumulation overhead by 47.9\%; (2) minimizing ADC quantization noise accumulation, resulting in small accuracy degradation—only 0.01\% for LeNet-5, 0.1\% for ResNet-18, 0.5\% for VGG-16, and 0.9\% for SNN.
Compared to vanilla convolution (vConv), CADC exhibits accuracy changes ranging from +0.11\% to +0.19\% for LeNet-5, -0.04\% to -0.27\% for ResNet-18, +0.99\% to +1.60\% for VGG-16, and -0.57\% to +1.32\% for SNN, across crossbar sizes from 64×64 to 256×256. Ultimately, a SRAM-based IMC implementation of CADC achieves 2.15 TOPS and 40.8 TOPS/W for ResNet-18 (4/2/4b), realizing a $11\times$–$18\times$ speedup and $1.9\times$–$22.9\times$ improvement in energy efficiency compared to existing IMC accelerators. 

\end{abstract}

\begin{IEEEkeywords}
Dendritic computing, Convolution, In-memory computing, Crossbar, Sparsity
\end{IEEEkeywords}

\section{Introduction} 
\begin{figure}[t]
\centerline{\includegraphics[width=\linewidth]{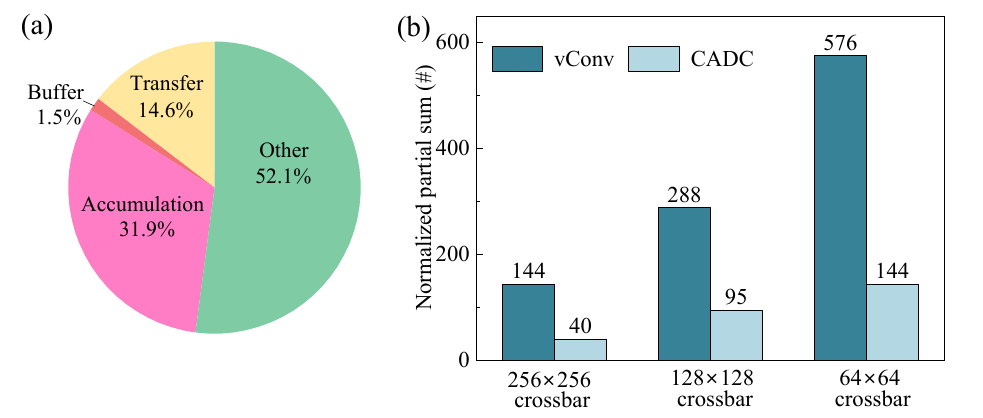}}
\caption{(a)  Energy breakdown of a 65 nm SRAM IMC accelerator running VGG-8 on CIFAR-10, as modeled with NeuroSim \cite{peng2020dnn+}. (b) Normalized count of psums comparison between vanilla convolution (vConv) and CADC when implementing 6-th convolution layer into different crossbars.}
\label{fig1:motivation}
\end{figure}

Convolution neural networks (CNNs) have become a fundamental architecture in various machine learning tasks. Due to their computational efficiency, In-memory computing (IMC) is a promising architecture to reduce memory bottlenecks by using crossbars for computation. However, as neural networks grow in size, particularly with large convolution layers, IMC systems face challenges due to the conflict between the increasing layer size and the limited physical size of crossbars. This necessitates partitioning large convolution layers, which generates large volumes of partial sums (psums) that must be buffered, transferred, and accumulated across multiple units, causing significant system overhead. As presented in Fig.~\ref{fig1:motivation}(a), psums contribute to 48\% of total energy consumption in the SRAM IMC-based VGG8 network evaluated on CIFAR-10, primarily due to (1) excessive buffer read/write operations associated with psums \cite{hu202528nm}, (2) frequent transfers\cite{li20241} of psums among crossbars, buffers and accumulators, and (3) increased quantity and complexity of accumulation operations \cite{hu202528nm} needed for the psums. To further explain this overhead, Fig.~\ref{fig1:motivation}(b) examines the sixth convolutional layer (with 8-bit weights) of the VGG8 network. Compared to the scenario without crossbar partitioning, the number of psums significantly increases by factors of approximately $144\times$ to $567\times$ when the convolution kernel is partitioned across $256\times256$, $128\times128$, and $64\times64$ SRAM IMC crossbars.

Several strategies have been used in software to mitigate psum overhead: 1) Quantization \cite{bai2023partial,saxena2023partial} reduces psum volume by using low-bit precision. 2) Structured pruning \cite{wang2021convolutional} eliminates redundant weight blocks, reducing the number of crossbars and psums. 3) Lightweight convolution, such as group \cite{wu2021software}, depthwise \cite{guo2019depthwise}, and factorized \cite{dong2023performance} convolutions, reduces kernel partitioning times. Although the above methods indeed cut down the cost of psums, they degrade accuracy and fail to address the problem at its root.
On the hardware side, an alternative approach involves designing larger crossbars which theoretically reduces the number of psums. For example, $512\times2048$ crossbars are fabricated in \cite{ambrogio2023analog}.  However, the manufacturing challenge is a significant obstacle, and the signal integrity problems (IR drop, sneak path) become more pronounced in large crossbars.   

To solve this problem, we look at neuroscience for inspiration. Biological neurons receive inputs on long dendrites which perform nonlinear operations on short weighted sum of inputs. Previous work has applied dendritic neurons to reduce synaptic resources, i.e. weight matrix size of shallow MLP by using sparse connection matrices \cite{acharya2022dendritic}. Recently, \cite{chavlis2025dendrites,zheng2024temporal,han2025manipulation} have shown similar results for slightly deeper layers but dendritic neurons are limited to the first layer alone. Also, no research so far has explored dendritic multi-layer CNNs and their hardware acceleration through IMC.

In this work, we introduce dendrites into convolution by applying a nonlinear function $f()$ to each crossbar's psums. This is based on the observation that each crossbar in a convolution layer captures part of the receptive field, drawing an analogy to biological dendrites that perform sparse and structured integration. To alleviate psums overhead, $f()$ is designed to clamp negative psums to zero. As shown in Fig.~\ref{fig1:motivation}(b), this approach reduces the number of psums by 72\%, 67\%, and 75\% in the three respective crossbar implementations. The resulting zeros enable efficient zero-compression and zero-skipping, significantly lowering buffer, transfer, and accumulation costs. We refer to this method as crossbar-aware dendritic convolution (CADC). 
The main contributions of this work are:
\begin{itemize}
  \item CADC: We introduce a novel CNN computation method that partitions convolutions across crossbars and applies nonlinear dendritic function $f()$ for each crossbar’s psums. This effectively reduces the psum numbers, eliminating 80\% in LeNet-5 on MNIST, 54\% in ResNet-18 on CIFAR-10, 66\% in VGG-16 on CIFAR-100, and up to 88\% in spiking neural networks (SNN) on DVS Gesture.
  
  \item CADC is evaluated across multiple benchmarks with varying crossbar sizes. It exhibits accuracy changes ranging from +0.11\% to +0.19\% for LeNet-5, -0.04\% to -0.27\% for ResNet-18, +0.99\% to +1.60\% for VGG-16, and -0.57\% to +1.32\% for SNN across 64×64, 128×128 and 256×256 crossbars.
  
  \item Four types of $f()$ are evaluated on CADC to model different dendritic behaviors. Among them, ReLU achieves highest accuracy for artificial neural networks (LeNet-5, Resnet-18 and VGG-16), while a sublinear function ($\sqrt x$) is more effective for SNN. Nonlinear in-memory ADC are used to implement $f()$ \cite{yang2025efficient}.

  \item The sparse psums generated by CADC alleviates ADC quantization noise accumulation effect, with only 0.01\%, 0.1\%, 0.5\%, and 0.9\% accuracy reduction due to noise for LeNet-5, ResNet-18, VGG-16, and SNN, respectively.
  \item Zero-compression and zero-skipping methods are used to eliminate 29.3\% buffer and transfer cost, and 47.9\% accumulation overhead. The simulated ResNet-18 (4/2/4b) on CIFAR-10 with CADC achieves $11\times-18\times$ faster and $1.9\times-22.9\times$ more energy efficient compared with existing SRAM IMC accelerators.
\end{itemize}

\begin{figure}[t]
\centerline{\includegraphics[width=\linewidth]{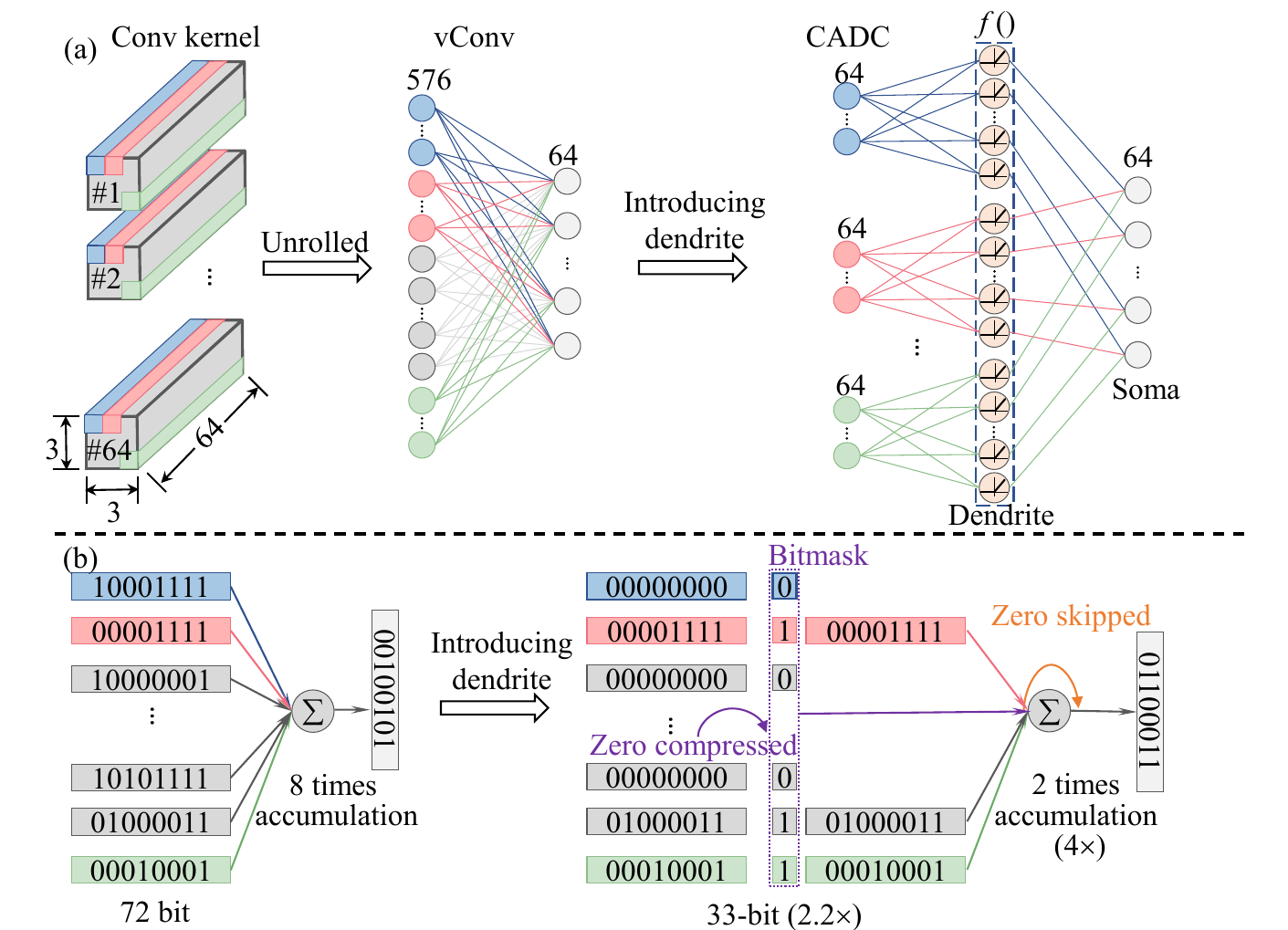}}
\caption{Comparison of CADC and vConv in (a) software and (b) hardware}
\label{fig:cadc methods}
\end{figure}

\section{Background}
\subsection{Dendritic computing}
Dendritic computing is inspired by the structure and function of biological neurons, where dendrites play a key role in the initial stages of signal processing by collecting electrical signals from synapses, which are then transmitted to the soma for further processing. The concept of dendritic computing has been applied to develop architectures that enable efficient computation by mimicking these biological processes. To show the difference between common neural networks and dendritic computing, we first give the equation about the former:
\begin{equation}
    y[k] = \sum_{i}w[i,k] \cdot x[i]
\end{equation}
where $x[i]$ is the input connect to $i$-th input neuron, $w[i,k]$ is the weight connecting $i$-th input neuron and $k$-th output neuron. $y[k]$ is the output of $k$-th output neuron. 

A simple mathematical model of dendritic computing is inserting dendritic nonlinearities between the input neuron and the output neuron (soma)\cite{chavlis2025dendrites}, which can be mathematically expressed as:
\begin{equation}
\label{eq:dendritic computing}
    y[k] = \sum_{j} w^{k}[j] \cdot f(\sum_{i}\left( w^{k,j}[i] \cdot x^{k,j}[i]\right))
\end{equation}
where $x^{k,j}$ represents the inputs to the $j$-th dendrite of the $k$-th soma, fetched by structured sampling from $x$. $x^{k,j}[i]$ denotes the $i$-th input to this dendrite. $w^{k,j}[i]$ is the corresponding weight connecting dendrite and input neuron. $w^{k}[j]$ is the weight connecting $j$-th dendrite to $k$-th soma. Biological experiments have shown the existence of both supralinear and sublinear forms of $f()$\cite{chavlis2025dendrites}.

\begin{figure}[t]
\centerline{\includegraphics[width=\linewidth]{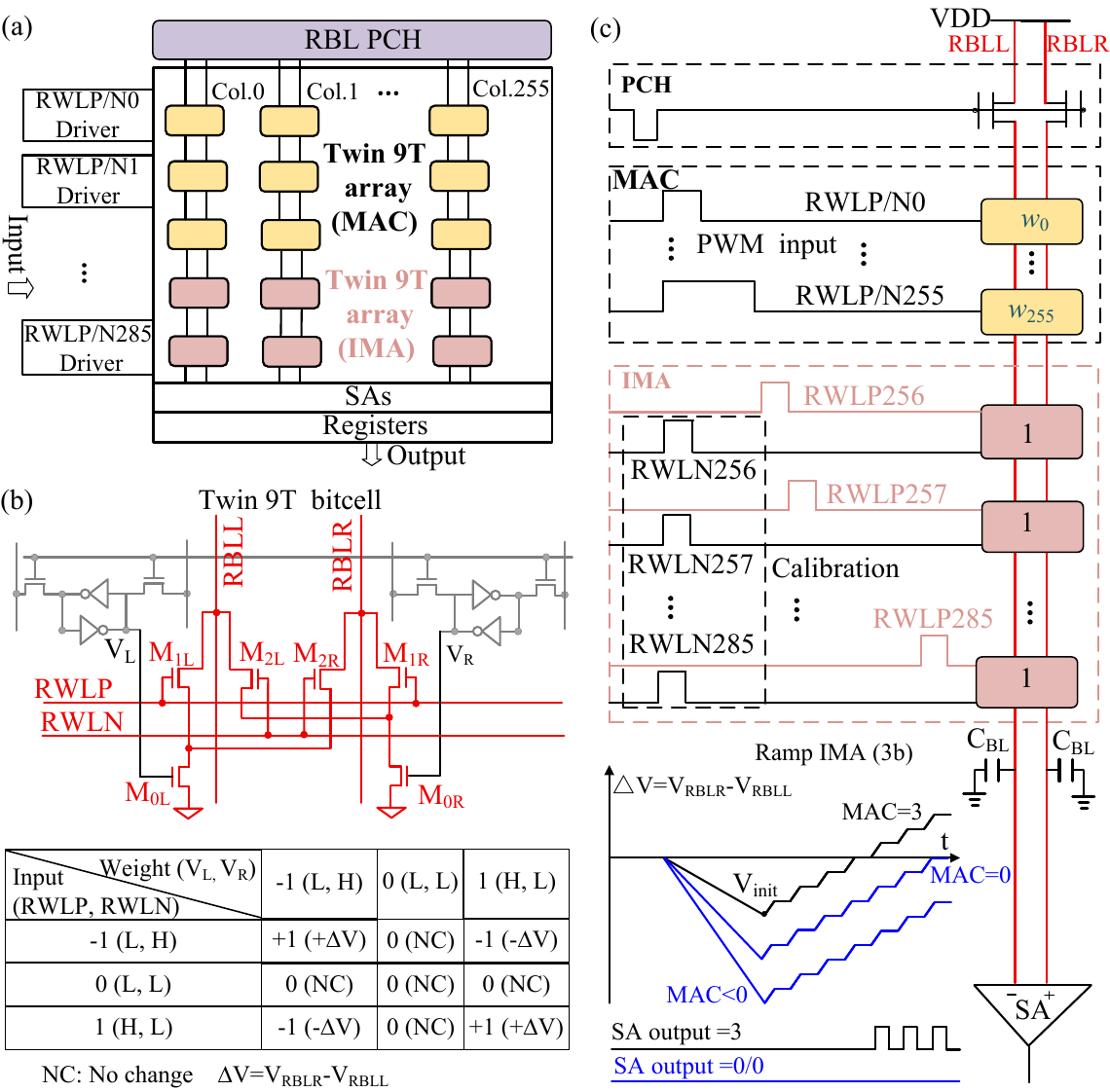}}
\caption{Overall macro structure: (a) Hardware structure. (b) Information of twin 9T SRAM bitcell. (c) Single-column circuit diagram with corresponding timing characteristics for MAC and IMA.}
\label{fig：chip_overall_architecture}
\end{figure}

\section{Design features}
To efficiently optimize the overhead of psums arising from the weight partitioning into a size-limited crossbar, we introduce the concept of dendritic computing into convolution. Additionally, we present the hardware implementation that integrates $f()$ within the crossbar.
\subsection{Crossbar aware dendritic convolution}
\label{sec:Crossbar aware dendritic convolution}
When implementing a convolution kernel of size $C_{in}\times K_1\times K_2\times C_{out}$ into a size-limited crossbar of size $N\times N$, the kernel is first unrolled along the input channel into a 2D network with dimensions $C_{in}\cdot K_1\cdot K_2 \times C_{out}$. However, the input dimension becomes prohibitively large due to the multiplication of these three factors($C_{in}$, $K_1$ and $K_2$). Therefore, a partitioning approach is required, wherein the input dimension is partitioned into $S=\lceil \frac{Cin\cdot K_1\cdot K_2}{N}\rceil$ segments. This partitioning results in $S\times$ psums. In vanilla convolution (vConv), all these psums must be accumulated to obtain the final output of the convolution layer. This accumulation can be expressed mathematically as:
\begin{equation}
\label{eq:Conv}
    y[k] = \sum_{s} \sum_{i}\left(  w^{s}[i,k] \cdot x^{s}[i]\right)
\end{equation}
where $w^{s}[i,k]$ represents the weight connecting the $i$-th input neuron of the $s$-th segment to the $k$-th output neuron. $x^{s}[i]$ is the corresponding unrolled input fetched by sliding the kernel over the input feature map. 
As shown in Fig.~\ref{fig:cadc methods} (a), a $64(C_{in})\times 3(K_1)\times 3(K_2)\times 64(C_{out})$ kernel is used as an example, requiring S=9 of $64\times64$ crossbars. In Fig.~\ref{fig:cadc methods}(b), we only consider the psums for one times convolution for one output neuron, where 9 psums are generated (72 bits in total, with each being 8 bits). These psums are buffered and transferred to the accumulation circuits, where they undergo an 8 times accumulation process.

To mitigate the computational overhead introduced by psums, dendritic computing is applied at the crossbar level, as illustrated in Fig.~\ref{fig:cadc methods} (a). Each crossbar focuses on $\frac{1}{S}$ of the local receptive field of the convolution kernel, which aligns with the sparse and structured connectivity behavior observed in dendrites. Drawing inspiration from this, CADC is proposed, which applies $f()$ to the psums from crossbars before accumulation. $f()$ is designed as $f(x)=0$ for $x\le 0$ and $f(x)=g(x)$ for $x>0$, where \( g \in \{\sqrt{x}\,\text{ (sublinear)},\; kx^{2}\,\text{ (supralinear)},\; \tanh (x),\; \operatorname{ReLU}(x)\}. \)
The results for these different $f()$ functions are presented in Sec. \ref{sec:sw result}.
$f()$ prunes negative psums to zero, thereby effectively reducing the computational overhead associated with psums. Inspired by Eq.~\ref{eq:dendritic computing}, the CADC operation can be mathematically expressed as:

\begin{equation}
\label{eq:CADC}
    y[k] = \sum_{s} w^{k}[s] \cdot f( \sum_{i}\left(  w^{s}[i,k] \cdot x^{s}[i]\right)
\end{equation}

where $w^{k}[s]$ is the weight connecting the $s$-th dendrite to the $k$-th soma. To minimize hardware overhead, $w^{k}[s]$ is set to 1. 

Using ReLU as an example in Fig.~\ref{fig:cadc methods} (b), the negative psums are clamped to zeros by $f()$. The resulting sparse set of nine psums is then compressed using a 9-bit bitmask alongside three 8-bit positive psums, substantially reducing both buffer storage and data transfer overhead. Consequently, the original psums are compressed by $2.2 \times$, resulting in a 33-bit data representation. Furthermore, zero psums are skipped in the accumulation circuits, reducing the required number of accumulations to two, leading to a $4\times$ improvement in efficiency.


\begin{figure}[t]
\centerline{\includegraphics[width=\linewidth]{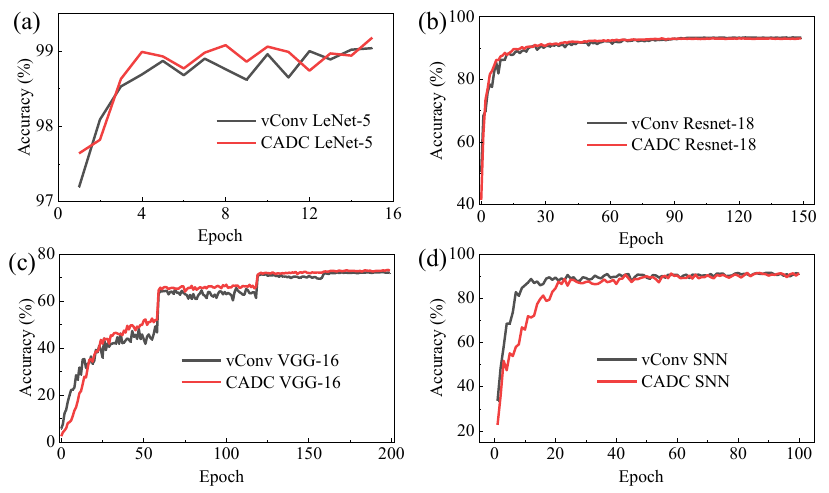}}
\caption{Comparative training analysis of CADC and vConv using LeNet-5 on MNIST, ResNet-18 on CIFAR-10, VGG-16 on CIFAR-100, and an SNN on the DVS Gesture dataset.}
\label{fig:training vs epoch}
\end{figure}

\begin{figure}[t]
\centerline{\includegraphics[width=\linewidth]{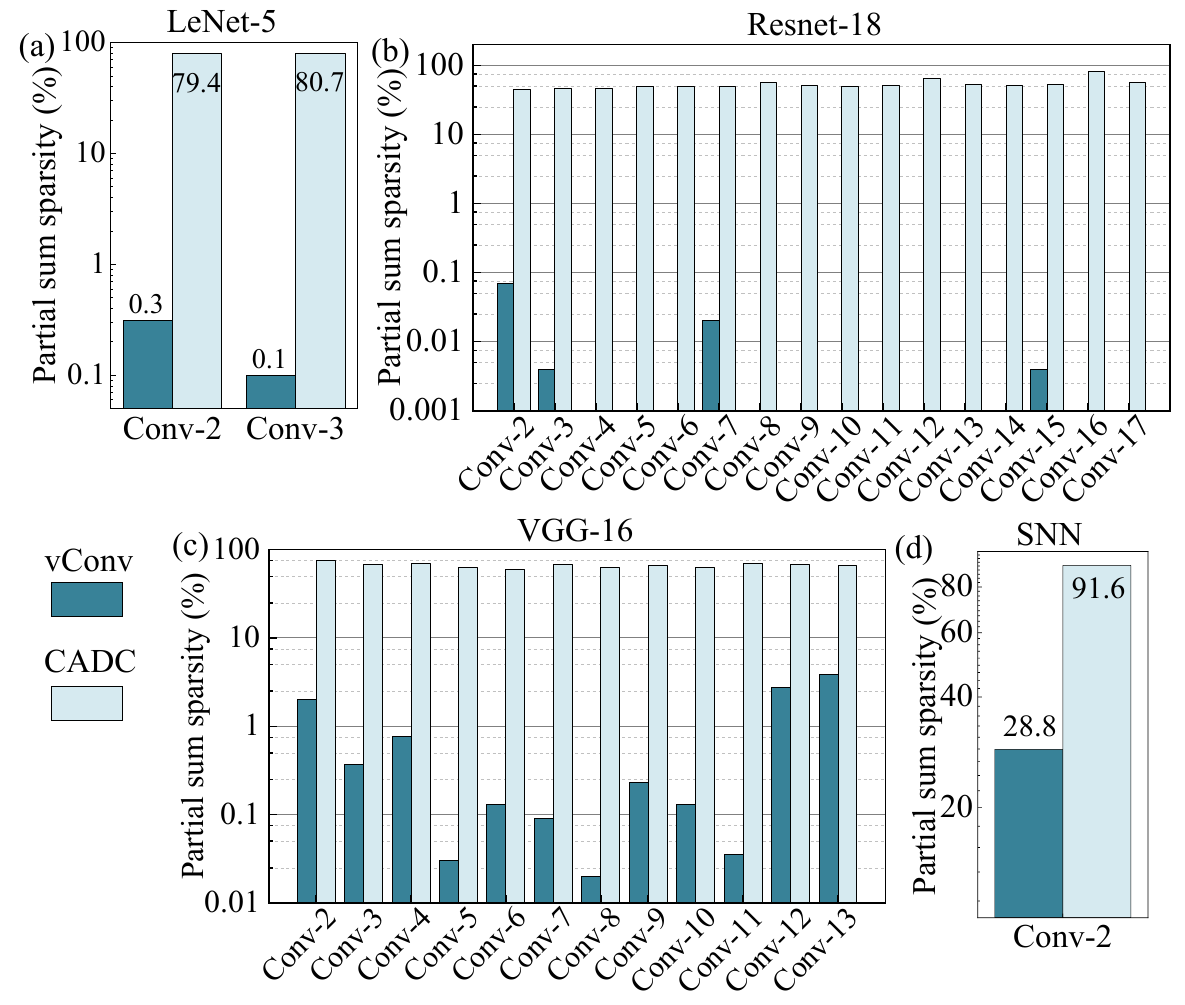}}
\caption{Psums sparsity comparison between vConV and CADC across all convolution layer in (a) LeNet-5 on MNIST, (b) Resnet-18 on CIFAR-10, (c) VGG-16 on CIFAR-100 and (d) SNN on DVS Gesture.} 
\label{fig:sparsity}
\end{figure}

\subsection{Hardware implementation of CADC}
To achieve the basic computation unit for CADC, an SRAM-based macro architecture is illustrated in Fig. \ref{fig：chip_overall_architecture}(a) that uses IMC for multiply-accumulate (MAC) and ADC operations (our proposed method is equally applicable to RRAM macros). The design comprises several key components: a computational crossbar of $256\times 256$ twin 9T SRAM bitcells for MAC operations, a $30\times 100$ bitcell array serving as reference cells for the reconfigurable (1-5 bit) In-memory ADC (IMA) to achieve dendritic function $f()$, along with peripheral circuits including read word line (RWL) buffers, sense amplifiers (SAs) and registers.

The macro is based on the twin 9T SRAM bitcell as shown in Fig. \ref{fig：chip_overall_architecture}(b) (top), which incorporates a decoupled read path consisting of six red NMOS transistors to enable ternary multiplication between signed inputs and ternary weights. The input polarity is determined by applying high voltage to either the positive read word line (RWLP) for positive inputs or the negative read word line (RWLN) for negative inputs, while the ternary weights are stored in the  6T-SRAM bitcell with three possible states: -1($V_{L}$=L, $V_{R}$=H); 0(($V_{L}$=L, $V_{R}$=L)) and +1(($V_{L}$=H, $V_{R}$=L)). When the stored weight is 0 ($V_L$=low, $V_R$=low), neither the left nor right read bit line (RBLL or RBLR) discharges regardless of input state, whereas for non-zero weights (-1 or +1), the corresponding read bit line discharges through the activated NMOS transistor whose gate is connected to the asserted RWL. The ternary multiplication result is represented by the voltage difference between the right and left RBLs ($\Delta V=V_{RBLR}-V_{RBLL}$) as shown in Fig. \ref{fig：chip_overall_architecture}(b), and the compact layout of this twin 9T bitcell measures 3.6 $\mu m$$\times$1.9 $\mu m$ using 65 nm process technology.

Fig. \ref{fig：chip_overall_architecture}(c) presents the timing diagrams for both MAC operations in pulse-width modulation (PWM) mode with multi-bit inputs and IMA operations. During the computation phase, PWM inputs for MAC operations and calibration inputs for IMA are applied concurrently, generating a differential voltage on two RBLs through current-mode operation. Subsequently, the IMA module generates a progressively decreasing reference voltage on RBLL ($V_{RBLL}$) at each clock cycle, effectively producing an increasing ramp voltage  as references of our IMA as shown in Fig. \ref{fig：chip_overall_architecture}(c) ($\Delta V$, bottom). 

Conventional IMA architectures \cite{yu202265} require additional $2^{n}$ ($n$ is the resolution of ADC) bitcells to generate the initial ramp voltage $V_{init}$ (Fig. \ref{fig：chip_overall_architecture}(c)). In contrast to conventional implementations, the proposed architecture eliminates these calibration bitcells through the innovative twin 9T bitcell design. This bitcell supports both positive (RWLP) and negative (RWLN) inputs, where negative inputs (RWLN256-RWLN285) serve to generate the $V_{init}$ while positive inputs (RWLP256-RWLP285) produce the ramp voltage signal as shown in the IMA implementation in Fig. \ref{fig：chip_overall_architecture}(c) (middle). This architectural enhancement significantly lowers the $V_{init}$ in Fig. \ref{fig：chip_overall_architecture}(c) (bottom) of the ramp reference signal. As a result, when the MAC output is less than or equal to zero (indicated by the blue curve in the Fig. \ref{fig：chip_overall_architecture}(c)), the SA consistently maintains zero outputs, thereby perfectly realizing the ReLU function. Beyond implementing ReLU, our IMA can also be configured into a nonlinear mode following the methodology presented in \cite{yang2025efficient}, thereby enabling the realization of various nonlinear functions (sublinear function($\sqrt{x}$), supralinear function ($kx^2$), $\tanh(x)$)  as discussed in section \ref{sec:Crossbar aware dendritic convolution}.

\begin{figure}[t]
\centerline{\includegraphics[width=\linewidth]{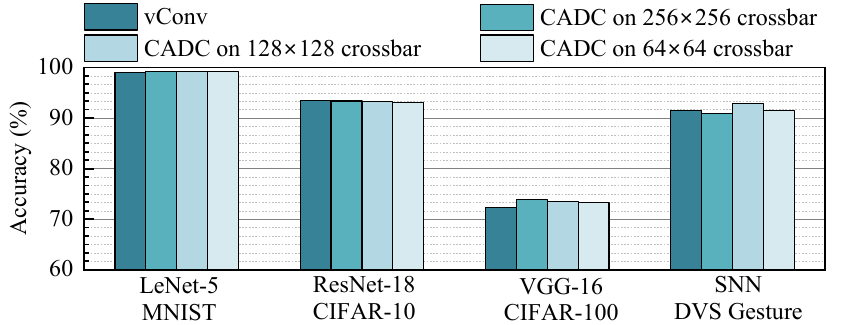}}
\caption{CADC evaluation among different crossbar sizes.}
\label{fig:accuracy vs crossbar_size}
\end{figure}

\begin{table}[t]
\begin{center}
\caption{Different nonlinear functions evaluation for CADC under $64\times64$ crossbar.}
\begin{tabular}{|l|c|c|c|c|}
\hline
                        & LeNet-5       & Resnet-18         & VGG-16            & SNN    \\ \hline
vConv (\%)               & 99.04         & 93.41             & 72.28             & 91.48  \\ \hline
CADC-ReLU (\%)  & $\textbf{99.18}$ &\textbf{ 93.14}     & \textbf{73.27}          & 91.48  \\ \hline
CADC-Sublinear(\%)      & 98.81         & 60.80             & 35.20    & \textbf{92.90}  \\ \hline
CADC-Supralinear(\%)    & 98.42         & 90.52             & 59.52                 & 91.67  \\ \hline
CADC-Tanh (\%) & 99.05         & 92.90             & 73.23             & 91.57  \\ \hline
\end{tabular}
\end{center}
\label{tab:nonlinear function}
\end{table}

\section{Simulation results}
\subsection{Software results}
\label{sec:sw result}
To assess the applicability of CADC, we conduct experiments on four models: LeNet-5 on MNIST, ResNet-18 on CIFAR-10, VGG-16 on CIFAR-100, and a SNN with two convolution layers and one fully connected layer on DVS Gesture. The baseline (vConv) accuracy for these models is 99.04\%, 93.41\%, 72.28\%, and 91.48\%, respectively.

\begin{figure*}[t]
\centerline{\includegraphics[width=0.95\linewidth]{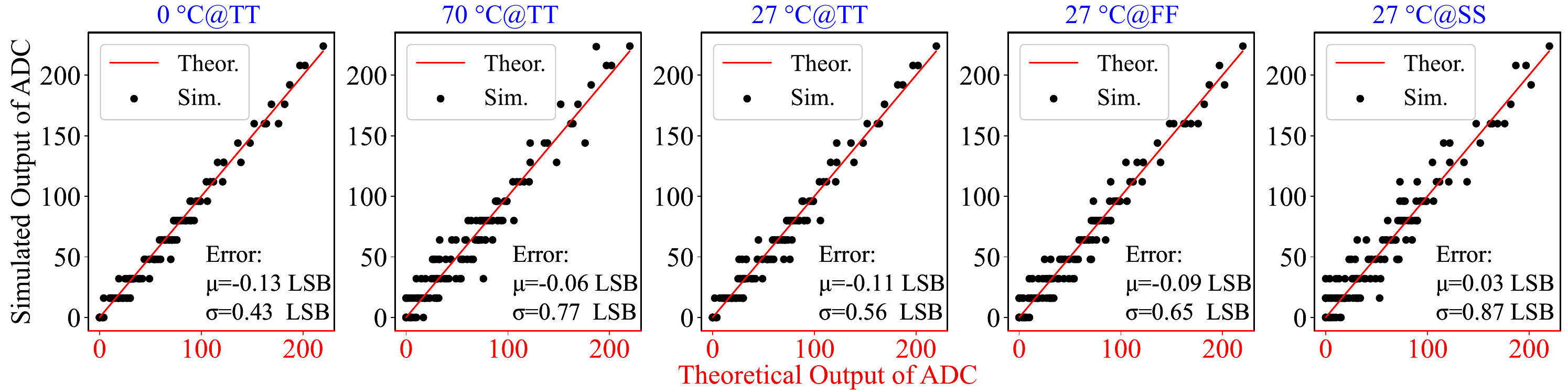}}
\caption{Simulated versus theoretical 4-bit ADC output across temperature variations (0°C, 70°C, 27°C) and different process corners (TT, FF, SS).}
\label{fig: temp_and_process}
\end{figure*}

\begin{figure}[t]
\centerline{\includegraphics[width=\linewidth]{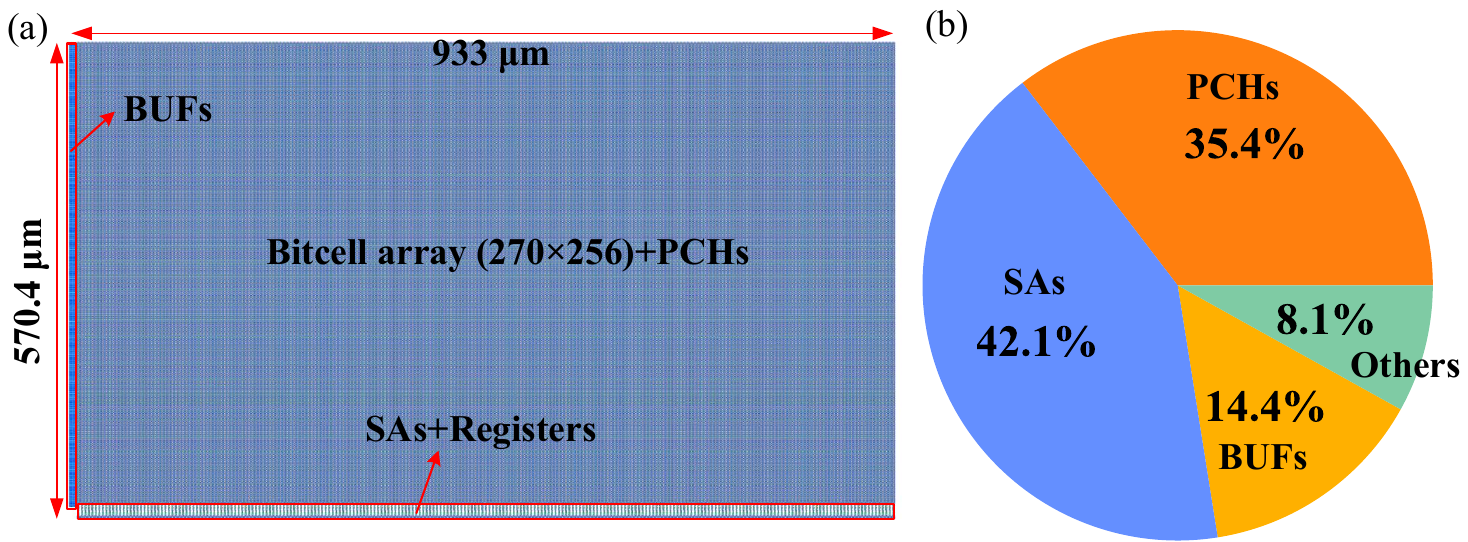}}
\caption{(a) Core layout of the macro (0.5 mm\textsuperscript{2}). (b)  Energy breakdown of the macro (4-bit input/output and 2-bit weight).}
\label{fig:Macro energy and area breakdown}
\end{figure}

As illustrated in Fig.~\ref{fig:training vs epoch}, the accuracy versus training epoch is presented for the $64 \times 64$ crossbar with the ReLU as $f()$. CADC demonstrates faster convergence compared to vConv for both LeNet-5 and VGG-16. It achieves comparable or slightly improved accuracy across these four models, with changes of +0.14\% (LeNet-5), -0.27\% (Resnet-18), +0.99\% (VGG-16), and 0\% (SNN), respectively. These results highlight the effectiveness of CADC in maintaining or slightly enhancing accuracy.

Owing to the role of $f()$ in CADC, negative psums are effectively eliminated, resulting in sparse psums. To further show the benefits of CADC, as shown in Fig.~\ref{fig:sparsity}, we analyze the sparsity of psums generated by vConv and CADC methods—across the convolution layers of four distinct models. Note that the first convolution kernel (Conv-1) is excluded in these four models, as it is small enough to be implemented on a single $64 \times 64$ crossbar, and therefore does not generate any psums. The sparsity of CADC across all layers ranges from 79.4\% to 80.7\% in LeNet-5, compared to 0.1\% to 0.3\% for vConv. In ResNet-18, CADC sparsity ranges from 45.1\% to 81.3\%, while vConv sparsity is from 0.0\% to 0.7\%. In VGG-16, CADC sparsity spans from 59.9\% to 74.6\%, with vConv ranging from 0.02\% to 3.83\%. In SNN, CADC sparsity is 91.6\%, whereas vConv is 28.8\%. This suggests that CADC effectively generates high sparsity in the psums, offering the potential to compress the sparse psums, thereby reducing buffer and transfer overhead. Additionally, it minimizes accumulation overhead by skipping zero psums.

After analyzing the psum sparsity, we evaluate the performance of CADC across different crossbar sizes to validate its broad applicability. As seen from Fig.~\ref{fig:accuracy vs crossbar_size}, CADC achieves highest accuracy under $128\times 128$ crossbar for LeNet-5, $256\times 256$ crossbar for Resnet-18, $256\times 256$ crossbar for VGG-16 and $128\times 128$ crossbar for SNN, with +0.19\%, -0.04\%, +1.6\% and +1.32\% accuracy changes compared with vConv.

The above result of CADC utilize ReLU as $f()$. To identify the most suitable nonlinear function for accurately modeling dendritic behavior in CADC across different tasks, we evaluate three additional $f()$. Table I shows that ReLU yields the highest accuracy for artificial neural networks (LeNet-5, ResNet-18, and VGG-16), whereas the sublinear function delivers the best performance for SNN.

\begin{figure}[t]
\centerline{\includegraphics[width=\linewidth]{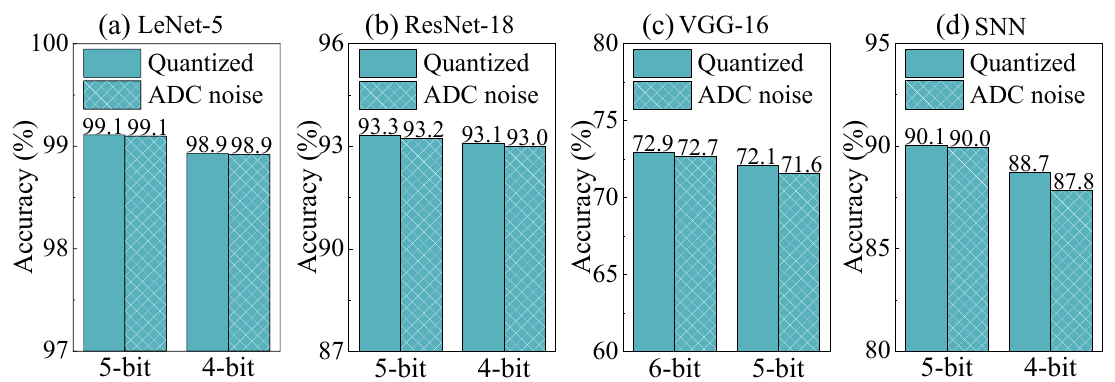}}
\caption{Quantization and test results in $27^\circ \text{C}$ for (a) LeNet-5, (b) Resnet-18, (c) VGG-16 and (d) SNN.}
\label{fig:quantization and noise}
\end{figure}

\subsection{Hardware results}
\textbf{CADC crossbar simulation.} SPICE simulations are performed in a 65 nm CMOS process to evaluate the proposed architecture. It utilizes dual clock domains: a 1 GHz clock for PWM inputs and a 62.5 MHz clock for IMA computations. The design operates with a 0.8 V pre-charge voltage and a 0.5 V supply voltage for the 6T-SRAM array. The simulated 4-bit  ADC outputs with 2-bit weight are compared against theoretical ADC outputs across varying temperatures ($0^\circ \text{C}$, $27^\circ \text{C}$, and $70^\circ \text{C}$) and process corners (TT, FF, SS) as illustrated in Fig. \ref{fig: temp_and_process}. The resulting errors exhibit an approximately normal distribution, with consistently low mean ($\mu$) and standard deviation ($\sigma$), demonstrating the robustness of the IMA against temperature and process variations because of replica biasing. 
Fig. \ref{fig:Macro energy and area breakdown}(a) illustrates the corresponding layout of CADC crossbar in 65 nm process, where the core occupies 0.5 mm². Notably, the 256 IMAs require merely 14.9\% of macro area, representing 1.5× and 3.8× area improvements compared to previous successive approximation register (SAR) ADC (21.7\%) \cite{zhang2023macc}  and IMA (57\%) \cite{yu202265}. Fig. \ref{fig:Macro energy and area breakdown}(b) presents the simulated energy breakdown of our macro for a 4-bit input/output and 2-bit weight configuration. The analysis reveals that pre-charging operations and SAs dominate the energy consumption. The proposed macro achieves an energy efficiency of 725.4 TOPS/W under these operating conditions.

\textbf{ADC quantization noise evaluation.} Then the nominal error distribution ($N(-0.11, 0.56)$ at $27^\circ \text{C}$) is used to inject noise into the four networks. As shown in Fig.~\ref{fig:quantization and noise}, the resulting accuracy loss is less than 0.01\%, 0.1\%, 0.5\%, and 0.9\% for the quantized LeNet-5 (4-5/2/4-5 bit), ResNet-18 (4-5/2/4-5 bit), VGG-16 (5-6/2/5-6 bit), and SNN (4-5/2/4-5 bit), respectively, benefiting from CADC-generated sparse psums that mitigate cumulative ADC quantization noise. Although the accumulation time for noisy psums is increased due to the larger kernel size of VGG-16 and the long time step of the SNN, CADC is still able to restore precision to a comparable accuracy. The highest ADC resolution used for these four models results in $0.05\%$, $0.14\%$ and $1.08\%$ and $0.94\%$ drop in accuracy from the floating point models based on CADC with $256\times256$ crossbars (99.15\%, 93.37\% and 73.88\% and 90.91\%), making them comparable to the vConv models (99.04\%, 93.41\%, 72.28\%, and 91.48\%). This indicates that the CADC enhances the models' robustness to noise.

\begin{figure}[t]
\centerline{\includegraphics[width=\linewidth]{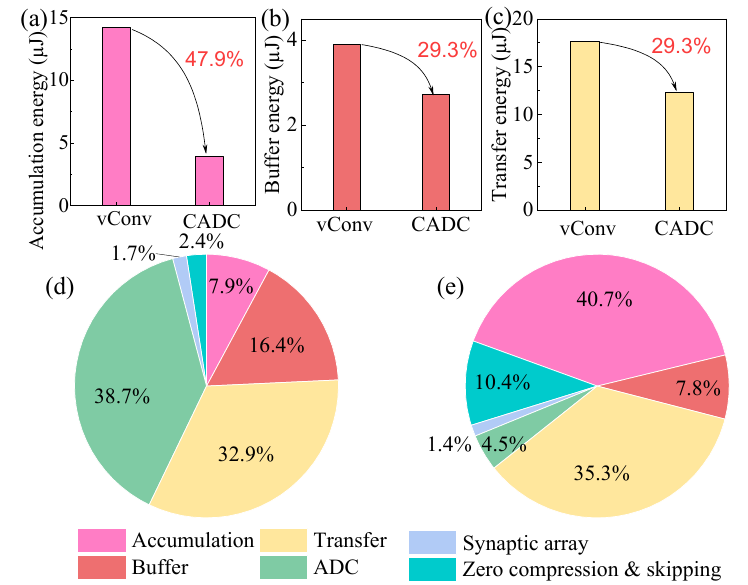}}
\caption{System evaluation for CADC Resnet-18 on CIFAR-10. (a)-(c) show the energy reductions in accumulation, buffer, and transfer achieved by CADC relative to vConv.(d) and (e) are the latency and energy breakdown results, receptively.}
\label{fig:performance breakdown}
\end{figure}

\textbf{System evaluation.} To demonstrate the benefits of the CADC within the entire system, ResNet-18 on the CIFAR-10 is evaluated. The overhead associated with buffer, transfer, and accumulation is simulated using NeuroSim \cite{peng2020dnn+} at 200 MHz. The costs for the crossbar and ADC is evaluated using SPICE simulations. Additionally, the zero-compression technique is adapted from \cite{kang20207}, wherein sparse psums are compressed into non-zero psums accompanied by a bitmask indicating the presence or absence of zero-valued psums. The zero-skipping approach is derived from \cite{kim2022distributed}. The overhead associated with these zero-compression and zero-skipping techniques is evaluated through digital synthesis in a 65 nm process at 200 MHz. As illustrated in Figs.~\ref{fig:performance breakdown}(a)–(c), CADC reduces accumulation energy by 47.9 \% because zero-skipping eliminates unnecessary accumulations of zero psums. Buffer and transfer energy drop 29.3 \% by zero compression. Figs.~\ref{fig:performance breakdown}(d) and (e) present the latency and energy breakdowns, respectively. The overhead of zero compression and skipping is relatively small but can significantly reduce the related psums overhead. Despite this reduction, the percentage of psums cost is still a little bit large mainly because the computation in crossbar is fast and energy-efficient in the IMC architecture. The advantages of CADC become more pronounced as psums sparsity increases and as designs demand higher-resolution ADCs. Even with a 4-bit ADC and 54\% sparsity, the CADC-enhanced ResNet-18 achieves up to $11\times$-$18\times$ speedup and $1.9\times$-$22.9\times$ more energy efficiency, compared to the existing SRAM IMC accelerators\cite{wang2023charge,chen20237,mao202428} in Table II.

\begin{table}[]

\begin{center}
\begin{threeparttable}
\caption{Comparison with state-of-the-arts}

\begin{tabular}{|c|c|c|c|c|}
\hline
& \begin{tabular}[c]{@{}c@{}}JSSC’22\\ \cite{yue2022sticker}\end{tabular}  
&\begin{tabular}[c]{@{}c@{}}ISSCC’23\\ \cite{chen20237}\end{tabular} 
&\begin{tabular}[c]{@{}c@{}}TCASI’24\\ \cite{mao202428} \end{tabular}
& Prop.\\ \hline
Tech. (nm)                          & 65        & 28           & 28               & 65   \\ \hline
Supp. (V)                           & 0.9-1.05       & 0.54-0.9     & 0.9-0.95         & 1.1  \\ \hline
Freq. (MHz)                         & 50-100    & 20-230       & 160-340          & 200\\ \hline
Bitcell                             & 8T        & 8T           & 9T1C             & Twin 9T  \\ \hline
Array size                          & 64×64     & 128×128      & 1152×128         &  256×256  \\ \hline      
I/O bit \#                          & 4/4       & 2-8/2-8      & 2,4,8/2,4,8      &  4/4    \\ \hline       
W. bit \#                           & 4,8       & 2-8          & 2,4,8            &  2    \\ \hline
Network                             & ResNet-18 & ResNet-20    & ResNet-18        & ResNet-18 \\ \hline
Dataset                             & CIFAR-10  & CIFAR-10     & CIFAR-10         & CIFAR-10 \\ \hline
Acc. loss (\%)\textsuperscript{1}   & 0.52      & 0.29         & 3.22             & 0.35 \\ \hline
TOPS                                & 0.20      & 0.12         & -                & 2.15    \\ \hline
TOPS/W\textsuperscript{2}           & 1.78-6.91 & 10.58        & 5.45-21.82      &  40.8    \\ \hline
\end{tabular}

    \begin{tablenotes}
      \item $^1$ Compared to software simulation of floating-point configuration  
      \item $^2$ TOPS/W=Reported TOPS/W$\times$ (Tech. /65nm)$\times$(Supp. /1.1V)$^2$

    \end{tablenotes}
  \end{threeparttable}
\end{center}
\label{tab1}
\end{table}

\section{Conclusion}
This paper introduces CADC, a dendritic computing-inspired method addressing computational overhead from partitioning large CNNs in crossbar-based IMC architectures. By embedding a zero-thresholding $f()$ directly within crossbar, CADC substantially reduces psums (80\% for LeNet-5, 54\% for ResNet-18, 66\% for VGG-16, and 88\% for SNNs). The introduced sparsity enables zero-compression and zero-skipping, reducing buffer/transfer overhead by 29.3\% and accumulation overhead by 47.9\%. Additionally, CADC significantly mitigates ADC-induced noise, resulting in minimal accuracy degradation ($<$0.9\%). Experimental validation shows comparable or superior accuracy relative to vConv, with performance variations of +0.11\% to +1.60\% and minor losses of -0.04\% to -0.57\%, across diverse CNN models and crossbar dimensions (64×64 to 256×256).
A SRAM-based IMC implementation of CADC delivers considerable gains, achieving 2.15 TOPS and 40.8 TOPS/W for ResNet-18 (4/2/4b), translating into 11×–18× speedup and 1.9×–22.9× energy efficiency improvement. The method is general and can be applied to RRAM IMC accelerators as well. 

\bibliography{IEEEref}

\begin{thebibliography}{10}
\providecommand{\url}[1]{#1}
\csname url@samestyle\endcsname
\providecommand{\newblock}{\relax}
\providecommand{\bibinfo}[2]{#2}
\providecommand{\BIBentrySTDinterwordspacing}{\spaceskip=0pt\relax}
\providecommand{\BIBentryALTinterwordstretchfactor}{4}
\providecommand{\BIBentryALTinterwordspacing}{\spaceskip=\fontdimen2\font plus
\BIBentryALTinterwordstretchfactor\fontdimen3\font minus \fontdimen4\font\relax}
\providecommand{\BIBforeignlanguage}[2]{{%
\expandafter\ifx\csname l@#1\endcsname\relax
\typeout{** WARNING: IEEEtran.bst: No hyphenation pattern has been}%
\typeout{** loaded for the language `#1'. Using the pattern for}%
\typeout{** the default language instead.}%
\else
\language=\csname l@#1\endcsname
\fi
#2}}
\providecommand{\BIBdecl}{\relax}
\BIBdecl

\bibitem{hu202528nm}
X.~Hu, H.~Mun, J.~Meng, Y.~Liao, A.~Sridharan, and J.-s. Seo, ``{A 28nm 20.9-137.2 TOPS/W Output-Stationary SRAM Compute-in-Memory Macro Featuring Dynamic Look-ahead Zero Weight Skipping and Runtime Partial Sum Quantization},'' in \emph{2025 IEEE Custom Integrated Circuits Conference (CICC)}.\hskip 1em plus 0.5em minus 0.4em\relax IEEE, 2025, pp. 1--3.

\bibitem{li20241}
D.~Li, M.~M. Wong, Y.~S. Chong, J.~Zhou, M.~Upadhyay, A.~Balaji, A.~Mani, W.~F. Wong, L.~S. Peh, A.~T. Do \emph{et~al.}, ``1.63 pj/sop neuromorphic processor with integrated partial sum routers for in-network computing,'' \emph{IEEE Transactions on Very Large Scale Integration (VLSI) Systems}, 2024.

\bibitem{bai2023partial}
J.~Bai, W.~Xue, Y.~Fan, S.~Sun, and W.~Kang, ``Partial sum quantization for computing-in-memory-based neural network accelerator,'' \emph{IEEE Transactions on Circuits and Systems II: Express Briefs}, vol.~70, no.~8, pp. 3049--3053, 2023.

\bibitem{saxena2023partial}
U.~Saxena and K.~Roy, ``Partial-sum quantization for near adc-less compute-in-memory accelerators,'' in \emph{2023 IEEE/ACM International Symposium on Low Power Electronics and Design (ISLPED)}.\hskip 1em plus 0.5em minus 0.4em\relax IEEE, 2023, pp. 1--6.

\bibitem{wang2021convolutional}
Z.~Wang, C.~Li, and X.~Wang, ``Convolutional neural network pruning with structural redundancy reduction,'' in \emph{Proceedings of the IEEE/CVF conference on computer vision and pattern recognition}, 2021, pp. 14\,913--14\,922.

\bibitem{wu2021software}
Q.~Wu, L.~Tao, H.~Liang, W.~Yuan, T.~Tian, S.~Xue, and X.~Jin, ``Software-hardware co-optimization on partial-sum problem for pim-based neural network accelerator,'' in \emph{2021 IEEE High Performance Extreme Computing Conference (HPEC)}.\hskip 1em plus 0.5em minus 0.4em\relax IEEE, 2021, pp. 1--7.

\bibitem{guo2019depthwise}
Y.~Guo, Y.~Li, L.~Wang, and T.~Rosing, ``Depthwise convolution is all you need for learning multiple visual domains,'' in \emph{Proceedings of the AAAI Conference on Artificial Intelligence}, vol.~33, no.~01, 2019, pp. 8368--8375.

\bibitem{dong2023performance}
S.~Dong, Z.~Fan, Y.~Chen, K.~Chen, M.~Qin, M.~Zeng, X.~Lu, G.~Zhou, X.~Gao, and J.-M. Liu, ``Performance estimation for the memristor-based computing-in-memory implementation of extremely factorized network for real-time and low-power semantic segmentation,'' \emph{Neural Networks}, vol. 160, pp. 202--215, 2023.

\bibitem{ambrogio2023analog}
S.~Ambrogio, P.~Narayanan, A.~Okazaki, A.~Fasoli, C.~Mackin, K.~Hosokawa, A.~Nomura, T.~Yasuda, A.~Chen, A.~Friz \emph{et~al.}, ``An analog-ai chip for energy-efficient speech recognition and transcription,'' \emph{Nature}, vol. 620, no. 7975, pp. 768--775, 2023.

\bibitem{acharya2022dendritic}
J.~Acharya, A.~Basu, R.~Legenstein, T.~Limbacher, P.~Poirazi, and X.~Wu, ``Dendritic computing: branching deeper into machine learning,'' \emph{Neuroscience}, vol. 489, pp. 275--289, 2022.

\bibitem{chavlis2025dendrites}
S.~Chavlis and P.~Poirazi, ``Dendrites endow artificial neural networks with accurate, robust and parameter-efficient learning,'' \emph{Nature Communications}, vol.~16, no.~1, p. 943, 2025.

\bibitem{zheng2024temporal}
H.~Zheng, Z.~Zheng, R.~Hu, B.~Xiao, Y.~Wu, F.~Yu, X.~Liu, G.~Li, and L.~Deng, ``Temporal dendritic heterogeneity incorporated with spiking neural networks for learning multi-timescale dynamics,'' \emph{Nature Communications}, vol.~15, no.~1, p. 277, 2024.

\bibitem{han2025manipulation}
X.~Han, S.~Mao, Y.~Wang, Y.~Lu, D.~Wang, Y.~Sun, Y.~Zheng, X.~Feng, L.~Lu, J.~Hua \emph{et~al.}, ``Manipulation of lithium dendrites based on electric field relaxation enabling safe and long-life lithium-ion batteries,'' \emph{Nature Communications}, vol.~16, no.~1, p. 3699, 2025.

\bibitem{peng2020dnn+}
X.~Peng, S.~Huang, H.~Jiang, A.~Lu, and S.~Yu, ``Dnn+ neurosim v2. 0: An end-to-end benchmarking framework for compute-in-memory accelerators for on-chip training,'' \emph{IEEE Transactions on Computer-Aided Design of Integrated Circuits and Systems}, vol.~40, no.~11, pp. 2306--2319, 2020.

\bibitem{yang2025efficient}
J.~Yang, R.~Mao, M.~Jiang, Y.~Cheng, P.-S.~V. Sun, S.~Dong, G.~Pedretti, X.~Sheng, J.~Ignowski, H.~Li \emph{et~al.}, ``Efficient nonlinear function approximation in analog resistive crossbars for recurrent neural networks,'' \emph{Nature Communications}, vol.~16, no.~1, p. 1136, 2025.

\bibitem{yu202265}
C.~Yu, T.~Yoo, K.~T.~C. Chai, T.~T.-H. Kim, and B.~Kim, ``{A 65-nm 8T SRAM compute-in-memory macro with column ADCs for processing neural networks},'' \emph{IEEE Journal of Solid-State Circuits}, vol.~57, no.~11, pp. 3466--3476, 2022.

\bibitem{zhang2023macc}
B.~Zhang, J.~Saikia, J.~Meng, D.~Wang, S.~Kwon, S.~Myung, H.~Kim, S.~J. Kim, J.-S. Seo, and M.~Seok, ``{MACC-SRAM: A multistep accumulation capacitor-coupling in-memory computing SRAM macro for deep convolutional neural networks},'' \emph{IEEE Journal of Solid-State Circuits}, vol.~59, no.~6, pp. 1938--1949, 2023.

\bibitem{kang20207}
S.~Kang, D.~Han, J.~Lee, D.~Im, S.~Kim, S.~Kim, and H.-J. Yoo, ``7.4 ganpu: A 135tflops/w multi-dnn training processor for gans with speculative dual-sparsity exploitation,'' in \emph{2020 IEEE International Solid-State Circuits Conference-(ISSCC)}.\hskip 1em plus 0.5em minus 0.4em\relax IEEE, 2020, pp. 140--142.

\bibitem{kim2022distributed}
D.~Kim and J.~Park, ``Distributed accumulation based energy efficient stt-mram based digital pim architecture,'' in \emph{2022 19th International SoC Design Conference (ISOCC)}.\hskip 1em plus 0.5em minus 0.4em\relax IEEE, 2022, pp. 29--30.

\bibitem{wang2023charge}
H.~Wang, R.~Liu, R.~Dorrance, D.~Dasalukunte, D.~Lake, and B.~Carlton, ``A charge domain sram compute-in-memory macro with c-2c ladder-based 8-bit mac unit in 22-nm finfet process for edge inference,'' \emph{IEEE Journal of Solid-State Circuits}, vol.~58, no.~4, pp. 1037--1050, 2023.

\bibitem{chen20237}
P.~Chen, M.~Wu, W.~Zhao, J.~Cui, Z.~Wang, Y.~Zhang, Q.~Wang, J.~Ru, L.~Shen, T.~Jia \emph{et~al.}, ``{7.8 A 22nm delta-sigma computing-in-memory ($\Delta\Sigma$ CIM) SRAM macro with near-zero-mean outputs and LSB-first ADCs achieving 21.38 TOPS/W for 8b-MAC edge AI processing},'' in \emph{2023 IEEE International Solid-State Circuits Conference (ISSCC)}.\hskip 1em plus 0.5em minus 0.4em\relax IEEE, 2023, pp. 140--142.

\bibitem{mao202428}
W.~Mao, D.~Liu, H.~Zhou, F.~Li, K.~Li, Q.~Wu, J.~Yang, Q.~Cheng, L.~Zhang, and H.~Yu, ``{A 28-nm 135.19 TOPS/W Bootstrapped-SRAM Compute-in-Memory Accelerator With Layer-Wise Precision and Sparsity},'' \emph{IEEE Transactions on Circuits and Systems I: Regular Papers}, 2024.

\bibitem{yue2022sticker}
J.~Yue, Y.~Liu, Z.~Yuan, X.~Feng, Y.~He, W.~Sun, Z.~Zhang, X.~Si, R.~Liu, Z.~Wang \emph{et~al.}, ``Sticker-im: A 65 nm computing-in-memory nn processor using block-wise sparsity optimization and inter/intra-macro data reuse,'' \emph{IEEE Journal of Solid-State Circuits}, vol.~57, no.~8, pp. 2560--2573, 2022.

\end{thebibliography}
\bibliographystyle{IEEEtran}

\end{document}